\newif\iffigs\figstrue

\documentclass[12pt]{article}
\iffigs
  \input epsf
\else
\fi

\batchmode
\newfont{\footscrfont}{rsfs10}
  \newfont{\footbbbfont}{msbm10}
  \newfont{\manfont}{manfnt}
\errorstopmode

\newif\ifscrf\scrftrue
\ifx\footscrfont\nullfont
  \scrffalse
\fi

\newif\ifamsf\amsftrue
\ifx\footbbbfont\nullfont
  \amsffalse
\fi

\def\ppnumber{\vbox{\baselineskip14pt\hbox{YITP-SB 01-08}
\hbox{hep-th/0102183}}}
\def\ppdate{February 2001}
\def\pplogo{\vbox{\kern-\headheight\kern -15pt
\halign{##&##\hfil\cr&{
\ppnumber}\cr\rule{0pt}{2.5ex}&\ppdate\cr}
}}

\makeatletter
\date{}
\def\dedicatory#1{\def\@date{\normalsize\it#1}}
\def\subjclass#1{\def\@thefnmark{}\@footnotetext{1991
    {\it Mathematics Subject Classification.} #1}}
\def\keywords#1{\def\@thefnmark{}\@footnotetext{
    {\it Key words and phrases.} #1}}

\def\ps@firstpage{\ps@empty \def\@oddhead{\hss\pplogo}%
  \let\@evenhead\@oddhead 
}
\def\maketitle{\par
 \begingroup
 \def\thefootnote{\fnsymbol{footnote}}
 \def\@makefnmark{\hbox
 to 0pt{$^{\@thefnmark}$\hss}}
 \if@twocolumn
 \twocolumn[\@maketitle]
 \else \newpage
 \global\@topnum\z@ \@maketitle \fi\thispagestyle{firstpage}\@thanks
 \endgroup
 \setcounter{footnote}{0}
 \let\maketitle\relax
 \let\@maketitle\relax
 \gdef\@thanks{}\gdef\@author{}\gdef\@title{}\let\thanks\relax}

\def\abstract{\if@twocolumn
\section*{Abstract}
\else \small
\begin{center}
{\bf ABSTRACT}
\end{center}
\quotation
\fi}

\def\thebibliography#1{\section*{References\@mkboth
 {REFERENCES}{REFERENCES}}\small\list
 {[\arabic{enumi}]}{\settowidth\labelwidth{[#1]}\leftmargin\labelwidth
 \advance\leftmargin\labelsep
 \usecounter{enumi}}
 \def\newblock{\hskip .11em plus .33em minus .07em}
 \sloppy\clubpenalty4000\widowpenalty4000
 \sfcode`\.=1000\relax}

\newif\iffn\fnfalse

\@ifundefined{reset@font}{\let\reset@font\empty}{} 
\long\def\@footnotetext#1{\insert\footins{\reset@font\footnotesize
    \interlinepenalty\interfootnotelinepenalty
    \splittopskip\footnotesep
    \splitmaxdepth \dp\strutbox \floatingpenalty \@MM
    \hsize\columnwidth \@parboxrestore
   \edef\@currentlabel{\csname p@footnote\endcsname\@thefnmark}\@makefntext
    {\rule{\z@}{\footnotesep}\ignorespaces
      \fntrue#1\fnfalse\strut}}}

\makeatother

\ifamsf
  \newfont{\bigbbbfont}{msbm10 scaled\magstep2}
  \newfont{\bbbfont}{msbm10 scaled\magstep1}  
  \newfont{\smallbbbfont}{msbm8}
  \newfont{\tinybbbfont}{msbm6}
  \newfont{\smallfootbbbfont}{msbm7}
  \newfont{\tinyfootbbbfont}{msbm5}
  \newfont{\biggthfont}{eufm10 scaled\magstep2}
  \newfont{\gthfont}{eufm10 scaled\magstep1}  
  \newfont{\smallgthfont}{eufm8}
  \newfont{\tinygthfont}{eufm6}
  \newfont{\footgthfont}{eufm10}
  \newfont{\smallfootgthfont}{eufm7}
  \newfont{\tinyfootgthfont}{eufm5}
\fi

\ifscrf
  \newfont{\scrfont}{rsfs10 scaled\magstep1}  
  \newfont{\smallscrfont}{rsfs7}
  \newfont{\tinyscrfont}{rsfs7}
  \newfont{\smallfootscrfont}{rsfs7}
  \newfont{\tinyfootscrfont}{rsfs7}
\fi

\ifamsf

\else
  \def\bigbbbfont{\bf}

\fi

\ifscrf
  \newcommand{\Scr}[1]{\iffn
    \mathchoice{\mbox{\footscrfont #1}}{\mbox{\footscrfont #1}}
    {\mbox{\smallfootscrfont #1}}{\mbox{\tinyfootscrfont #1}}\else
    \mathchoice{\mbox{\scrfont #1}}{\mbox{\scrfont #1}}
    {\mbox{\smallscrfont #1}}{\mbox{\tinyscrfont #1}}\fi}
\else
  \def\Scr{\cal}
\fi

\def\bearray{\begin{eqnarray}}
\def\eearray{\end{eqnarray}}
\def\bearraynn{\begin{eqnarray*}}
\def\eearraynn{\end{eqnarray*}}
\def\bfig{\begin{figure}}
\def\efig{\end{figure}}

\def\opeq#1{\advance\lineskip#1 \advance\baselineskip#1
        \advance\lineskiplimit#1}

\def\cM{{\Scr M}}

\def\cD{{\Scr D}}

\def\cMc{{\hfuzz=100cm\hbox to 0pt{$\;\overline{\phantom{X}}$}\cM}}
\def\barcD{{\hfuzz=100cm\hbox to 0pt{$\;\overline{\phantom{X}}$}\cD}}

\ifamsf

\else

\fi

\def\boldone{\relax{\rm 1\kern-.35em 1}}

\newtheorem{Proposition}{Proposition}[section]

\newtheorem{Theorem}{Theorem}[section]
\newtheorem{Lemma}{Lemma}[section]
\newtheorem{Corrolary}{Corrolary}[section]

\newcommand{\be}{\begin{equation}}
\newcommand{\ee}{\end{equation}}
\newcommand{\bea}{\begin{eqnarray}}
\newcommand{\eea}{\end{eqnarray}}

\newcommand{\bp}{\begin{Proposition}}
\newcommand{\ep}{\end{Proposition}}
\newcommand{\bt}{\begin{Theorem}}
\newcommand{\et}{\end{Theorem}}
\newcommand{\bl}{\begin{Lemma}}
\newcommand{\el}{\end{Lemma}}
\newcommand{\bc}{\begin{Corrolary}}
\newcommand{\ec}{\end{Corrolary}}
\newcommand{\nn}{\nonumber}

\usepackage{graphics}

\begin{document}

\title{Unitarity, D-brane dynamics and D-brane categories}

\author{C.~I.~Lazaroiu$^{*}$}

\date{}

\maketitle

\vbox{ \centerline{C.~N.~Yang Institute for Theoretical Physics} 
\centerline{SUNY at Stony Brook} 
\centerline{NY11794-3840, U.S.A.} 
\medskip 
\medskip
\bigskip }

\abstract{This is a short nontechnical 
note summarizing the motivation and results of 
my recent work on D-brane categories. I also give a brief outline of 
how this framework can be applied to study the dynamics 
of topological D-branes and why this has a bearing on the homological mirror 
symmetry conjecture.  
This note can be read without any knowledge of category theory.
}

\vskip .6in

$*$ calin@insti.physics.sunysb.edu

\pagebreak

\tableofcontents

\pagebreak

\section{Introduction}

At the heart of the second superstring revolution one finds a duality 
in our description of D-brane dynamics. On one hand, D-branes are introduced 
at the fundamental level as boundary conditions in open string theory, while 
on the other hand string dualities together with the M-theory interpretation 
force us to treat them as dynamical objects. There is considerable 
fuzz surrounding the passage from `Dirichlet boundary conditions' to 
`dynamical objects'. In its most standard incarnation, 
the argument given takes the following indirect form. 

Starting with Dirichlet boundary conditions at the fundamental level, one 
obtains new open string sectors associated with strings ending on the brane. 
One next considers the low energy effective action of such {\em strings}, 
and identifies it with an effective description of low energy 
{\em D-brane} kinematics (the DBI action coupled to background fields). 
This gives us a low energy description of string fluctuations around the 
D-brane, and not a description of interactions between D-branes, hence our 
use of the term {\em kinematics}. 

The DBI action is obviously insufficient for a 
description of low energy D-brane {\em dynamics}. 
Indeed, the effective action of open strings ending on a D-brane 
describes the low energy dynamics of {\em strings} with 
prescribed boundary conditions, but the boundary condition itself is not  
`dynamical'  in any fundamental way. To describe D-brane interactions, 
one can 
resort to studies of string exchange between D-branes, consider the 
resulting low energy effective action and treat it as an effective description 
of D-brane interactions (this of course won't give anything interesting 
for collections of mutually BPS D-branes in type II theories, but there is 
no reason to restrict to type II or BPS saturated D-branes). Then one 
can study D-brane interactions through the dynamics of this 
action. However, effective 
actions do not give a fundamental (microscopic) description, 
and the way in which `boundary conditions' become dynamical is hard to 
see from such considerations. What, then, is D-brane dynamics ?

A conceptual approach to this issue is afforded by open string field 
theory. This allows one to answer some dynamical
questions at a fundamental level, as demonstrated explicitly by studies of 
tachyon condensation \cite{tachyon}. In fact, open string 
tachyon condensation is perhaps 
the only known example of {\em true}\footnote{By true dynamics we mean 
processes involving interaction and decay of branes, 
which in particular involve 
`second quantization'. 
In this language, 
the oscillations of a given D-brane would correspond to its `first 
quantization'.}
 D-brane dynamics described in a 
microscopic manner. Through such a process, D-branes 
are allowed to annihilate, decay, or form 
bound states. In a certain sense, passage to string field theory 
performs their `second quantization'. 

There are a few obvious lessons to be learned from studies of tachyon 
condensation. First, a truly dynamical description of D-branes requires
second quantization of strings and off-shell techniques, i.e. string field 
theory. Second, the notion of D-brane has to be extended. 

The second point follows from the observation that the end product of a 
condensation process is generally 
not a Dirichlet brane, since it typically cannot be described 
through boundary conditions on a worldsheet theory. For 
example, tachyon condensation in superstring compactifications on 
Calabi-Yau manifolds can produce D-brane composites described by 
various configurations of bundles and maps\cite{Oz_triples}, 
for which a direct worldsheet description through a boundary 
condition is not always available\footnote{Passage to the derived 
category as in \cite{Douglas_Kontsevich} allows for a representation of 
{\em some} such objects as coherent sheaves, some of which can in turn be 
identified with bundles living on complex submanifolds. However, not every 
object of the derived category is a coherent sheaf, and not every coherent 
sheaf can be represented in the second way.}. This implies 
that, at least in geometrically nontrivial backgrounds,  
tachyon condensation processes can produce genuinely new objects, 
distinct from the Dirichlet branes originally considered in the theory. 

Moreover, 
consideration of various condensates in a given background shows that they 
will generally interact through string exchange. It follows that such 
condensates behave in many respects as `abstract D-branes', even though they 
do not admit a direct description through boundary conditions. This 
implies that open string theory must be generalized to allow for a description 
of such objects.

One is thus lead to the task of formulating open string field theory in the 
presence of `abstract D-branes'. Since these are not simply boundary conditions, 
one has to find a structure which allows for their systematic description. 
The main point of \cite{com1} is that the correct structure (at least for the 
`associative case' ) is a so-called 
{\em dG (or differential graded) category}. This mathematical object 
arises naturally from constructions based on Dirichlet branes, 
and -- in a slightly less
direct manner -- also in the case of generalized D-branes (D-brane composites). 
Moreover, it is showed in \cite{com1} that D-brane composite formation can be 
described as a change of this structure. We are lead to the following:

\

{\bf Proposal} In first nontrivial approximation, D-brane dynamics is 
described by certain deformations of a dG category.

\

By first nontrivial approximation we mean the fact that we only consider tree 
level dynamics of open strings. Moreover, this approach treats all Dirichlet 
branes and their condensates on a equal footing (`bootstrap'), though it also 
opens the way for finding a `system of generators' which need not be 
of Dirichlet type.
The work of \cite{com1}, which I shortly review below, is concerned with  
formulating and exploring some basic consequences of this proposal.

\section{dG categories on one leg}

I now give a short description of the dG category describing usual (i.e. 
Dirichlet) D-branes. A category \cite{MacLane} 
is a collection of objects $a$ and 
sets $Hom(a,b)$ associated to any ordered pair of objects $a,b$, together with 
compositions $(u,v)\rightarrow uv$ for elements $u$ of $Hom(b,c)$ and 
$v$ of $Hom(a,b)$, where $uv$ belongs to $Hom(a,c)$. 
Such compositions are required to be associative, i.e. 
$(uv)w=u(vw)$, and to admit units $1_a$ (elements of the sets $Hom(a,a)$ )
such that $u1_a=1_bv$ for $u, v$ in $Hom(a,b)$.
The objects $a$ and `morphism sets' $Hom(a,b)$ can be pretty much 
anything as long as these requirements are satisfied. 
Familiar examples are the 
category ${\bf Ens}$ 
of sets (with the morphism space between two sets $A,B$ given by 
all functions from $A$ to $B$ and morphism compositions given by composition 
of 
functions), the category of vector spaces ${\bf Vct}$ 
(with morphisms given by linear maps),
and the category ${\bf Vect}(X)$ of 
vector bundles over a manifold $X$ (with morphisms 
given by bundle morphisms). In all of these cases the elements are some sets 
(with extra structure) and the morphisms are maps between these sets which 
preserve the structure  (these are so-called `concrete' categories). 
A category need not be of this type, however: its objects may not be sets, and 
its morphisms need not be maps of sets. 

As it turns out, Dirichlet branes in an 
associative oriented 
open string theory give an example of a (generally non-concrete) 
category ${\cal A}$. This arises by taking Dirichlet 
branes as objects and the morphism space 
between two objects to be given by the off-shell state space of open strings 
stretched between them (figure 1). 
In general, this space contains the full tower of 
massive modes, and therefore such morphisms cannot be naturally 
thought of as maps. 
The composition of morphisms is given by the string product of 
\cite{Witten_SFT}, which is related to the triple correlator on the disk 
(figure 1). In an associative string theory, this product is associative 
off-shell\footnote{In more general situations, the product 
need only be associative 
up to homotopy; this leads to an $A_\infty$ category upon extending the 
structure discussed in \cite{Gaberdiel} (see also \cite{Zwiebach_open}) to 
the case of backgrounds containing D-branes.}. 
Moreover, one has units 
$1_a$, related to the boundary vacua of \cite{top}; these are generalizations 
of the formal unit of cubic string field theory.

\iffigs
\hskip 0.8 in
\begin{center} 
\scalebox{0.6}{\begin{picture}(0,0)%
\includegraphics{cat.pstex}%
\end{picture}%
\setlength{\unitlength}{4144sp}%
\begingroup\makeatletter\ifx\SetFigFont\undefined%
\gdef\SetFigFont#1#2#3#4#5{%
  \reset@font\fontsize{#1}{#2pt}%
  \fontfamily{#3}\fontseries{#4}\fontshape{#5}%
  \selectfont}%
\fi\endgroup%
\begin{picture}(7349,2715)(1126,-1816)
\put(1126,-1636){\makebox(0,0)[lb]{\smash{\SetFigFont{17}{20.4}{\familydefault}{\mddefault}{\updefault}
$a$
}}}
\put(2791,-1591){\makebox(0,0)[lb]{\smash{\SetFigFont{17}{20.4}{\familydefault}{\mddefault}{\updefault}
$b$
}}}
\put(1981,-151){\makebox(0,0)[lb]{\smash{\SetFigFont{17}{20.4}{\familydefault}{\mddefault}{\updefault}
$\psi$
}}}
\put(6661,-1456){\makebox(0,0)[lb]{\smash{\SetFigFont{17}{20.4}{\familydefault}{\mddefault}{\updefault}
$c$
}}}
\put(7336,-241){\makebox(0,0)[lb]{\smash{\SetFigFont{17}{20.4}{\familydefault}{\mddefault}{\updefault}
$b$
}}}
\put(5896,-331){\makebox(0,0)[lb]{\smash{\SetFigFont{17}{20.4}{\familydefault}{\mddefault}{\updefault}
$a$
}}}
\put(8236,-1456){\makebox(0,0)[lb]{\smash{\SetFigFont{17}{20.4}{\familydefault}{\mddefault}{\updefault}
$u$
}}}
\put(5041,-1816){\makebox(0,0)[lb]{\smash{\SetFigFont{17}{20.4}{\familydefault}{\mddefault}{\updefault}
$uv$
}}}
\put(6616,659){\makebox(0,0)[lb]{\smash{\SetFigFont{17}{20.4}{\familydefault}{\mddefault}{\updefault}
$v$
}}}
\end{picture}
}
\end{center}
\begin{center} 
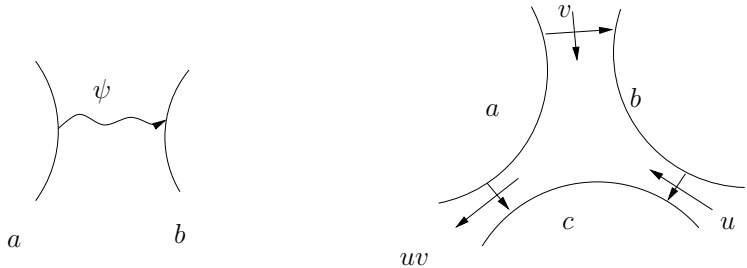
Figure  1. {\footnotesize Dirichlet branes define a category whose morphisms 
are off-shell states of oriented open strings stretched between them (left). 
Compositions of morphisms are given by the string product, which is related to 
the triple correlator (right).}
\end{center}
\fi

As it happens, the resulting category of Dirichlet branes has some extra 
structure which reflects the basic data of open string field theory. First, 
the off-shell state spaces $Hom(a,b)$ are graded by the ghost degree
\footnote{In a topological A/B string theory, this is replaced by the 
anomalous $U(1)$ charge of the twisted superconformal algebra.}. If one uses 
appropriate conventions for the ghost charge, the composition of morphisms 
preserves this degree, in the sense that the degree of $uv$ is the sum of 
degrees of $u$ and $v$. In technical language, this means that we have a 
{\em graded category}. Another essential ingredient is the worldsheet BRST 
charge, which defines linear operators $Q_{ab}$ on each of the spaces 
$Hom(a,b)$; as in \cite{Witten_SFT}, these operators act as derivations 
of the string product. With our conventions, they also have ghost degree
$+1$. A graded category endowed with degree one nilpotent
\footnote{Remember that an operator is nilpotent if it squares to zero.} 
operators on its 
Hom spaces, acting as derivations of morphism compositions, is known 
as a {\em differential graded} (dG, for short) category\cite{Keller_dg, 
Bondal_Kapranov}. It follows
that the Dirichlet branes of any associative string theory form a dG category.
In fact, a complete specification of open string field theory requires 
some more data, for example a collection of bilinear pairings on morphisms 
and possibly some complex conjugation operations, which are required to 
satisfy certain properties. I shall neglect this extra structure in order 
to simplify the discussion; the bilinear forms are treated in detail in 
\cite{com1}. 

\section{D-brane processes as shifts of the string vacuum}

We saw above that Dirichlet branes form a dG category. Does this structure 
also describe backgrounds containing D-brane condensates ? As we shall 
see in a moment, the answer is affirmative, though the dG category arising 
after formation of D-brane composites does not admit a direct construction 
in terms of string worldsheets (in fact, its description requires consideration 
of off-shell string dynamics). The nontrivial 
fact that a dG category can be used to 
describe both Dirichlet branes and generalized branes arising from 
condensation of boundary operators is what allows us to view the dG 
category structure as fundamental. 

The basic idea behind this approach is that D-brane composites represent new 
{\em boundary sectors}. To understand this, note that a background containing 
Dirichlet branes can also be described in terms of a `total 
boundary state space':
\be
\label{foo}
{\cal H}=\oplus_{a,b}{Hom(a,b)}~~,
\ee
endowed with the multiplication induced by morphism compositions. In this 
approach, one is given a $dG$ algebra, i.e. a vector space ${\cal H}$ endowed 
with an associative multiplication and a linear operator 
$Q=\oplus_{a,b}{Q_{ab}}$,
which  squares to zero and acts as a derivation of the product. This is 
precisely the algebraic framework of \cite{Witten_SFT}, expressed with our 
conventions for the ghost grading. The new input represented by the Dirichlet 
branes can be described as a decomposition property of the product. 
Namely, we have a decomposition (\ref{foo}) of ${\cal H}$ which has the 
property that the string product vanishes on subspaces of the form 
$Hom(b',c)\times Hom(a,b)$ unless $b'=b$, in which case it maps 
$Hom(b,c)\times Hom(a,b)$ into $Hom(a,c)$. This decomposition also has 
the property that it is preserved by the total BRST operator $Q$, i.e. 
$Q$ preserves each boundary sector $Hom(a,b)$.

At least formally, a decomposition 
of ${\cal H}$ having these properties is the only piece of data 
distinguishing the open string field theory of \cite{Witten_CS} from 
a theory containing various Dirichlet branes. The component spaces 
${\cal H}_{ba}=Hom(a,b)$ of such a decomposition will be called {\em boundary 
sectors}. Hence {\em 
the underlying D-brane category is determined by the properties 
of the total string product and total BRST charge}.

This point of view allows us to recover a category structure after formation 
of D-brane composites takes place. Indeed, such processes are described by 
condensation of certain boundary/boundary condition changing operators, which 
correspond to various states $q_{ab}$ 
in the boundary sectors $Hom(a,b)$. From the 
point of view of string field theory, this amounts to giving VEVs $q_{ab}$ 
to various components $\phi_{ab}\in Hom(a,b)$ 
of the total string field $\phi=\oplus_{a,b}{\phi_{ab}}\in {\cal H}$. 
It follows that the result of a condensation process can be described by the 
standard device of shifting the string vacuum. Such a shift 
$\phi\rightarrow \phi+q$ preserves the total boundary product, but 
induces a new BRST operator $Q'$. Moreover, the condition that 
the new vacuum extremizes the string field action 
imposes constraints on the allowed shifts $q$. The important 
observation is that the BRST operator $Q'$ for the shifted vacuum will generally
fail to preserve the original boundary sectors $Hom(a,b)$; this signals the 
fact that the collection of D-branes in the shifted background has changed, 
which is exactly what one expects from formation of D-brane composites. One can 
identify the new boundary sectors (and thus the composite D-branes and the 
state spaces they determine) by looking for a new decomposition of ${\cal H}$ 
into subspaces, which has the required compatibility properties with 
respect to the modified BRST operator $Q'$ and the boundary product.
This analysis is carried out 
in \cite{com1}, with the conclusion that the resulting boundary sectors 
form a new dG category ${\cal A}_q$, the so-called {\em contraction} of the 
original category ${\cal A}$ along the collection of boundary operators 
$q$ (figure 2). 
The objects and morphism spaces of this category are given explicitly 
in \cite{com1}. 

The conclusion is that D-brane composites can once again 
be described in terms of a dG category, even though they do not generally 
correspond to Dirichlet branes. Moreover, D-brane composite formation can be 
described as a change of the dG category structure. This justifies our proposal 
that the fundamental objects of interest are not Dirichlet branes per se, but 
rather abstract dG category structures. 
This amounts to generalizing D-branes to abstract boundary sectors, the latter 
being specified by decomposition properties of the total boundary product and 
BRST charge. As discussed above, this generalization is unavoidable if one 
wishes 
to allow for D-brane condensation processes, i.e. describe D-branes as truly 
dynamical objects. 

\
\iffigs
\hskip 1.0 in
\begin{center} 
\scalebox{0.6}{\begin{picture}(0,0)%
\epsfbox{cont.pstex}%
\end{picture}%
\setlength{\unitlength}{4144sp}%
\begingroup\makeatletter\ifx\SetFigFont\undefined%
\gdef\SetFigFont#1#2#3#4#5{%
  \reset@font\fontsize{#1}{#2pt}%
  \fontfamily{#3}\fontseries{#4}\fontshape{#5}%
  \selectfont}%
\fi\endgroup%
\begin{picture}(7238,2468)(432,-2150)
\end{picture}
}
\end{center}
\begin{center} 
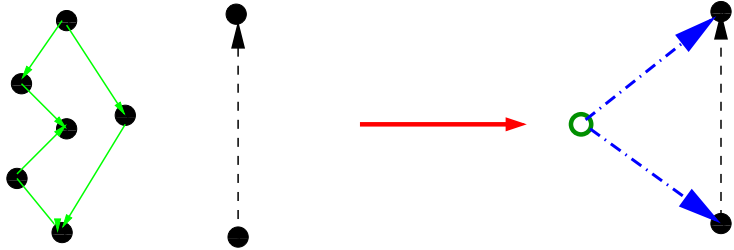
Figure  2. {\footnotesize Condensation of boundary condition changing 
operators leads to a new category structure. On the left, the green 
full lines represent boundary operators which acquire vevs, while the dashed 
black line is an operator which does not condense. 
The contracted category 
on the right is obtained by collapsing the objects 
connected by the operators which acquire vevs 
to a single object and building the associated morphism spaces and 
BRST charges 
in  a systematic manner which is explained in \cite{com1}. The new object 
(hollow circle) on 
the right represents the D-brane composite formed by condensation of the 
boundary condition changing operators. Some morphisms 
between this object and another object of the contracted category are 
represented by blue dash-and-dot lines. The dashed vertical line represents 
a morphism which does not change}
\end{center}
\fi

\section{Unitarity}

A basic condition on any description of dynamics is that the underlying space 
of states be closed under dynamical processes. Since condensation of 
boundary condition changing operators between Dirichlet branes 
leads to objects which generally cannot be described through Dirichlet boundary 
conditions, an open string theory whose boundary sectors are described by 
Dirichlet branes will typically fail to 
give a unitary description of D-brane dynamics: its 
boundary space must be enlarged. 

Does a suitable enlargement always exist ? As shown in \cite{com1}, the answer 
to this question is affirmative. More precisely, it can be argued that a 
`minimal' description which is closed under formation of D-brane composites 
can be obtained by enlarging the Dirichlet brane category ${\cal A}$ to 
its so-called {\em quasiunitary cover} $c({\cal A})$. This follows from 
careful consideration of successive extensions of the original category by 
adding condensates of Dirichlet branes, condensates of such condensates and 
so on. The category $c({\cal A})$ can be built explicitly as a category of 
generalized complexes over ${\cal A}$, i.e. sequences of objects
$(a_i)$ of ${\cal A}$ indexed by some finite set $I$, together with morphisms 
$q_{ij}\in Hom^1(a_i,a_j)$ of ghost degree one. These morphisms are subject 
to the condition:
\be
Q_{a_ia_j}q_{ij}+\sum_{k\in I}{q_{kj}q_{ik}}=0~~,
\ee
which is a generalized form of the string equations of motion. 

Unitarity of this description 
follows from the seemingly simple fact that we allow for {\em sequences} 
of objects $a_i$, and a sequence may have repetitions. This means that some 
of the Dirichlet branes  $a_i$ may coincide. It is important to realize that 
we are not talking about deformations of a given D-brane (such as its 
translations) but literally about identical D-branes in a given sequence. 
This is illustrated in figure 3.

\iffigs
\hskip 0.8 in
\begin{center} 
\scalebox{0.5}{\begin{picture}(0,0)%
\includegraphics{complex.pstex}%
\end{picture}%
\setlength{\unitlength}{4144sp}%
\begingroup\makeatletter\ifx\SetFigFont\undefined%
\gdef\SetFigFont#1#2#3#4#5{%
  \reset@font\fontsize{#1}{#2pt}%
  \fontfamily{#3}\fontseries{#4}\fontshape{#5}%
  \selectfont}%
\fi\endgroup%
\begin{picture}(4872,4980)(394,-4516)
\put(721,209){\makebox(0,0)[lb]{\smash{\SetFigFont{17}{20.4}{\familydefault}{\mddefault}{\updefault}
$a_1$
}}}
\put(1936,254){\makebox(0,0)[lb]{\smash{\SetFigFont{17}{20.4}{\familydefault}{\mddefault}{\updefault}
$a_2$
}}}
\put(3106,209){\makebox(0,0)[lb]{\smash{\SetFigFont{17}{20.4}{\familydefault}{\mddefault}{\updefault}
$a_3$
}}}
\put(2296,-1366){\makebox(0,0)[lb]{\smash{\SetFigFont{17}{20.4}{\familydefault}{\mddefault}{\updefault}
$a_4$
}}}
\put(1261,-2536){\makebox(0,0)[lb]{\smash{\SetFigFont{17}{20.4}{\familydefault}{\mddefault}{\updefault}
$q_{14}$
}}}
\put(1261,-3166){\makebox(0,0)[lb]{\smash{\SetFigFont{17}{20.4}{\familydefault}{\mddefault}{\updefault}
$q_{34}$
}}}
\put(496,-4111){\makebox(0,0)[lb]{\smash{\SetFigFont{17}{20.4}{\familydefault}{\mddefault}{\updefault}
$q_{23}$
}}}
\put(721,-4516){\makebox(0,0)[lb]{\smash{\SetFigFont{17}{20.4}{\familydefault}{\mddefault}{\updefault}
$a_1=a_2=a_3$
}}}
\put(946,-1006){\makebox(0,0)[lb]{\smash{\SetFigFont{17}{20.4}{\familydefault}{\mddefault}{\updefault}
$q_{14}$
}}}
\put(2386,-961){\makebox(0,0)[lb]{\smash{\SetFigFont{17}{20.4}{\familydefault}{\mddefault}{\updefault}
$q_{34}$
}}}
\put(1081,-196){\makebox(0,0)[lb]{\smash{\SetFigFont{17}{20.4}{\familydefault}{\mddefault}{\updefault}
$q_{12}$
}}}
\put(2296,-196){\makebox(0,0)[lb]{\smash{\SetFigFont{17}{20.4}{\familydefault}{\mddefault}{\updefault}
$q_{23}$
}}}
\put(2071,-3436){\makebox(0,0)[lb]{\smash{\SetFigFont{17}{20.4}{\familydefault}{\mddefault}{\updefault}
$a_4$
}}}
\put(541,-1636){\makebox(0,0)[lb]{\smash{\SetFigFont{17}{20.4}{\familydefault}{\mddefault}{\updefault}
$q_{12}$
}}}
\put(5266,-331){\makebox(0,0)[lb]{\smash{\SetFigFont{20}{24.0}{\familydefault}{\mddefault}{\updefault}
(1)
}}}
\put(5266,-3031){\makebox(0,0)[lb]{\smash{\SetFigFont{20}{24.0}{\familydefault}{\mddefault}{\updefault}
(2)
}}}
\end{picture}
}
\end{center}
\begin{center} 
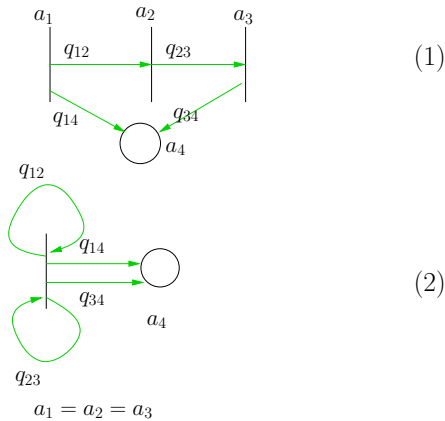
Figure  3. {\footnotesize Two generalized complexes with four terms. 
The D-branes $a_1$, $a_2$ and $a_3$ of the first complex are {\em distinct} 
objects even though they are translates of each other. That is, D-branes are 
{\em  not} identified if they differ by a translation (why should they ?); 
two parallel and non-coincident D-branes are treated as distinct. 
The second complex contains three identical terms $a_1$, $a_2$ and $a_3$; 
these correspond to the {\em same} D-brane $a$. The morphisms $q_{12}$ and 
$q_{23}$ of the second complex correspond to boundary  
operators belonging to the sector $Hom(a,a)$. One can formally 
think of the second complex as involving repeated 
condensation of the D-brane $a$ `with itself'. This is not the same 
as condensation of a D-brane with some of its translates, which is what is 
involved in the first complex.
}
\end{center}
\fi

It is easy to mistake this result for the trivial statement that repetitions 
in a generalized complex   
simply amount to condensation of a D-brane with its deformations (such as its 
translates (if such translates exist in the given background), 
as shown in part (1) of the figure). This 
is {\em not} what we mean, and I hope that the example of figure 3 
does something to prevent misunderstanding. `Condensation of a D-brane 
with itself' should be viewed as a way of performing deformations of the 
D-brane. For example, one 
could have a brane wrapping a special Lagrangian cycle in a type II 
Calabi-Yau compactification and condense 
a gauge field in order to produce a flat connection on the cycle. Then one 
could repeat the process by condensing around this new gauge background. 
This can be viewed as moving in the moduli space of flat connections on the 
cycle, i.e. as performing deformations of the associated D-brane. Finally, 
the end product of a few such deformations could form a composite with another 
D-brane (this would correspond to the second complex of figure 3, except that 
all D-branes should be viewed as wrapped).
One could 
similarly condense the lowest component of a chiral superfield 
living in the normal bundle, which amounts 
to deforming the cycle itself. Note, 
however, that this is a low energy description of such deformations, and in 
a physically realistic theory one would have to condense higher mass string 
modes as well, in order to satisfy the string field equations of motion. This 
gives a notion of `stringy D-brane deformations' (moduli) 
which takes into account all $\alpha'$ (and possibly worldsheet instanton) 
corrections, without making recourse to a description  
through effective actions. 
Though figure 3 suggests 
a flat background and noncompact D-branes, this is for reasons of 
simplicity only\footnote{As a matter of fact, there are good reasons 
(coming from the analysis of the A-twisted topological string theory) 
to believe that A-type D-branes in a type II compactification give a 
nonassociative category (likely an $A_\infty$ category), 
due to worldsheet instanton 
effects. In the simplified discussion above we have neglected this subtlety.}.

Despite such intuitive examples, 
it should be clear that there is 
no trivial {\em universal} justification for introducing 
generalized complexes, since condensing 
a D-brane with itself multiple times is harder to 
understand in, say, a string background based on an abstract 
conformal 
field theory\footnote{There are other reasons why it is inadvisable 
to base our arguments on low energy intuition, among them the fact that 
this would make it difficult to give a conceptual proof of unitarity.}. 
Moreover, our intuitive discussion above is based on low energy 
considerations, 
which can 
be modified rather markedly by stringy effects. 

Luckily, a completely general 
motivation for considering such objects is given by the procedure of 
recursively including condensates, condensates of condensates and so on in 
order to obtain a category which is closed under formation of composites. 
The fact that generalized complexes
(together with natural morphisms and morphism compositions) form a dG category 
which is closed under D-brane composite formation is proved in \cite{com1}. 

Finally, I should probably mention that, though I sometimes freely use 
the term `D-brane condensates', I make no claim as to the stability of 
such composites, which is inaccessible to the  simple moduli space analysis
employed in \cite{com1}. Our composites may in fact be stable, metastable or 
unstable (=`excited states'), but this has no bearing on the unitarity 
constraint which requires that they all be included in a complete description 
of the theory. Also note that working with the quasiunitary cover $c({\cal A})$
is similar to a `bootstrap' approach. 
What is a `minimal' set of generators, and what are its properties, is 
a question for which I do not presently have a good general 
answer, though I will 
endeavor to  propose some speculations in the conclusions of this note. 

\section{The topological case, twisted complexes and the derived category} 

I now wish to argue that applying these ideas to the 
topological B/A models \cite{Witten_NLSM, Witten_mirror, Witten_CS}
has something to do with derived 
categories and homological mirror symmetry\cite{Kontsevich}
(this is developed in more detail in \cite{com3}).
The first step toward making 
this connection is that Dirichlet branes in a topological string theory are 
graded objects in a natural manner, a fact which can be related to 
{\em extended} deformations of the string vacuum. 
This means that any Dirichlet 
brane $a$ has `formal translates' 
$a[n]$ for each integer $n$, such that $a[0]=a$. 
For the topological B model, the objects $a[n]$ can be identified with the 
graded bundles of \cite{Douglas_Kontsevich}(whose grading is given by a 
choice of branch for the argument of  the associated central charge 
\cite{pi_stab}). It can be shown that consideration 
of extended vacuum deformations requires an enlargement of a topological 
D-brane category ${\cal A}$ 
to its so-called {\em shift completion} ${\tilde {\cal A}}$. This is 
a dG category 
with objects $a[n]$ and with a dG structure induced from that of ${\cal A}$. 
Its morphism spaces are given by:
\be
Hom(a[n],b[m])=Hom(a,b)[m-n]~~,
\ee
i.e. result by shifts of the grading on the space $Hom(a,b)$ by $m-n$. 
It is clear that  
degree one boundary operators of the shift-completed theory correspond  
to boundary operators of arbitrary degree between the trivially graded 
D-branes $a=a[0]$. This captures the intuition that in a topological string 
theory it should be possible to condense boundary operators of an arbitrary 
degree, since the grading on the state spaces $Hom(a,b)$ is somewhat 
conventional\footnote{While the degree of the bosonic string field must always 
be one in our conventions (which follows from the fact \cite{Witten_SFT} 
that states of non-unitary degree are spurious), such a constraint 
is not physically fundamental 
in a topological string theory, whose grading is induced by the anomalous 
$U(1)$ symmetry of the twisted $N=2$ superconformal algebra and can therefore be 
shifted.}. Condensation of such operators provides the link with extended 
vacuum deformations. 

When combined with our previous results, this observation implies that a 
topological D-brane category ${\cal A}$ should be extended to the quasiunitary 
cover 
${\cal B}=c({\tilde {\cal A}})$ of its shift completion ${\tilde {\cal A}}$. 
The objects of ${\cal B}$ are so-called 
{\em generalized twisted complexes} over 
${\cal A}$, i.e. sequences of objects of ${\cal A}$ together with 
morphisms of `arbitrary' degree between them which satisfy a certain  
version of the string field equations of motion. These objects 
are closely related to those considered in \cite{Bondal_Kapranov}. 
Particular examples are the standard complexes:
\be
\label{boo}
a_1\stackrel{f_{12}}{\rightarrow} a_2 \stackrel{f_{23}}{\rightarrow} 
a_3...\stackrel{f_{n-1n}}{\rightarrow}a_n~~, 
\ee
with $f_{i,i+1}$ some degree {\em zero} morphisms in $Hom(a_i, a_{i+1})$.
In this situation, the underlying constraint on morphisms reduces to
the conditions that each $f_{ii+1}$ be BRST closed and that (\ref{boo}) is 
a complex in the category ${\cal A}$:
\bea
\label{moo}
Q_{a_ia_{i+1}}f_{ii+1}&=&0~~\nn\\
f_{i+1i+2}f_{ii+1}&=&0~~.
\eea
In the case of Calabi-Yau topological B models, whose trivially graded 
Dirichlet branes $a=a[0]$ 
are given by holomorphic vector bundles, 
objects of the type (\ref{boo}) 
reduce to complexes of homolorphic vector bundles 
and holomorphic maps. Indeed, the morphism space between two 
bundles $E$ and $F$ 
(i.e. the off-shell state space of strings stretched from $E$ to $F$) 
is in this case given by the space $\Omega^{0,*}(E^*\otimes F)$ of 
forms of types $(0,q)$ valued in the bundle $E^*\otimes F$. The degree of such 
a form is given by $q$. Hence degree zero states between trivially graded 
D-branes correspond to (smooth) bundle maps from $E$ to $F$. On the other 
hand, the BRST charge is given by the Dolbeault operator ${\overline \partial}$ 
coupled to $E^*\otimes F$ and hence conditions (\ref{moo}) reduce to the 
requirements that the maps $f$ are holomorphic and that they form a complex.

This recovers from a string field theory perspective the complexes considered in 
\cite{Douglas_Kontsevich,Oz_triples}. 
Note, however, that our approach does not 
make use of brane/antibrane pairs (since we work directly in the 
topological B model) and therefore is not subject to some of the 
limitations affecting the arguments of the papers just cited. 
In the approach of those 
papers, one works with the full type II superstring theory instead of the 
topological B model (which only describes the chiral primary sector of the 
former). As a 
consequence, one has a well-defined notion of antibrane and one identifies 
our degree zero boundary operators as tachyons. In order to be able to 
do this, 
however, one must consider sequences whose consecutive elements form
brane-antibrane pairs. This is not necessary in our approach, since we view 
condensation of boundary operators as a dynamical process in 
{\em topological} 
string theory. In fact, one can recognize the similarity between some ideas 
of \cite{Douglas_Kontsevich} and our string field arguments, though our 
approach differs through our insistence of working {\em off-shell} 
as much as possible. By contrast, the approach of \cite{Douglas_Kontsevich}
involves introducing the derived category at an early stage through the 
use of heuristic arguments. In our view, assumptions on D-brane dynamics 
(which in a hidden form underlie the evidence for the derived category 
presented in \cite{Douglas_Kontsevich}) should be {\em derived} from open 
string field theory. It is unclear if such a  dynamical analysis can 
be carried out at present in superstring field theory, since 
its formulation seems to be incompletely understood. This is why we 
retreat to the topological B string, which satisfies the associative framework 
of \cite{com1}.

Moreover, we obtain 
considerably more objects than the standard complexes (\ref{boo}). 
This seems to imply some extension of homological mirror symmetry from the 
derived category to a certain enlargement. Indeed, it can be shown that,  
if one starts with the category ${\cal A}$ whose elements are 
holomorphic bundles, and whose morphisms are the spaces  
$\Omega^{0,*}(E^*\otimes F)$,
then the quasiunitary cover of ${\tilde {\cal A}}$ is large enough to 
recover the derived category $D^b(X)$ of the underlying Calabi-Yau manifold 
$X$, upon performing appropriate localization 
with respect to quasi-isomorphisms. However, it seems that one
can obtain more than $D^b(X)$, which suggests that the original proposal 
of \cite{Kontsevich} could be extended to a larger category on the B model 
side. 

A similar discussion can be carried out for the A-model, leading to 
an enlargementff of the category originally considered in \cite{Kontsevich}. 
In this case, the situation is complicated by the fact that worldsheet 
instanton effects 
violate off-shell associativity of the string product, which implies that 
the underlying string field theory is described by an $A_\infty$ category
\cite{Fukaya, Fukaya2, Keller}.
This is further compounded by the fact that the finite radius 
string vacuum does not satisfy the string equations of motion
\cite{Fukaya2, boundary}, which implies 
that the vacuum must be shifted away from its large radius limit. It is 
possible to generalize our unitarity arguments to the $A_\infty$ case and 
arrive at a notion of 
quasiunitary cover of the shift completion of an $A_\infty$ 
category, which turns out to represent a certain off-shell variant 
of a proposal made in \cite{Kontsevich}. 

Since A-model non-associativity is purely a worldsheet instanton effect, 
it is not present at large radius and one can describe 
this limit in more standard mathematical terms. At a large 
radius point, one in fact has an associative string field theory which 
fits into the framework of \cite{com1}. This theory must be extended
to its shift cmpletion as discussed above, which leads to topological D-branes 
as graded objects. The latter can be identified with the graded Lagrangian 
submanifolds of \cite{Seidel}. The quasiunitary cover of the shift completion 
can once again be described rather concretely. 

\section{Conclusions and speculations}

The essence our approach is 
that it allows for a systematic description of D-brane dynamics by 
exploiting the crucial string field theoretic insight that a correct 
understanding of string physics requires an off-shell analysis. We believe 
that a systematic study along these lines can shed light on many crucial
issues in D-brane physics such as the relevance of K-theory
\cite{Witten_K, Witten_K_review, Moore_top, K_bos} and 
the classification of D-brane
composites. On the other hand, I would like to propose the 
{\em problem of generators}\footnote{I thank Radu Roiban for 
a useful conversation on this subject.}, namely to find a system of generators of the
quasiunitary cover $c({\cal A})$ which is minimal in an appropriate sense. 
More precisely, given a quasiunitary theory ${\cal B}$, 
one would like to find a category ${\cal A}_m$ such that 
$c({\cal A}_m)={\cal B}$ 
and such that ${\cal A}_m$ does not admit a strict subcategory with this 
property. Intuitively, a minimal system of generators would allow one to 
think of all objects of ${\cal B}$ as composites of the objects of 
${\cal A}_m$, much in the same way that baryons and mesons are composites 
of quarks and gluons in QCD. In other words, we are asking if there 
is some good notion of `elementary branes'.  It is not clear whether 
a minimal system of generators would be unique in an appropriate sense, 
and to what extent it would be background 
dependent. Note also that such 
generators need not be Dirichlet branes (i.e. need not admit a description 
through Dirichlet boundary conditions). Finding such minimal generating 
sets may be the best hope to understand the (typically very large) object 
$c({\cal A})$ without performing the set of daunting computations that 
seem to be implied in carrying out our approach for 
a nontopological string theory.  
For the topological case 
of Calabi-Yau B-models, a promising suggestion is offered by the 
theory of so-called helices (see for example  \cite{helices}). 
In this regard, I should perhaps also mention a tantalizing 
similarity with Matrix theory. Providing a good system of generators 
essentially amounts to describing all Dirichlet branes and their condensates 
as composites of some particular objects. From this perspective, Matrix 
theory amounts to the proposal that D-paticles may suffice, but this now 
seems to be insufficient in view of difficulties to reconstruct the 
full spectrum on geometrically nontrivial backgrounds. 
In our approach, this may be surmounted by the fact that a minimal system of 
generators may have to include a larger/different class of objects in order 
to achieve the generation property and some measure of background independence.

The approach outlined above may shed some 
light on the problem of relating closed and open string moduli. 
This is intimately connected with the suggestion made in \cite{Hofman, top} 
that a {\em unitary} formulation of open-closed 
string theory may lead to a certain 
equivalence between the open and closed sectors, as suggested 
by the physical interpretation \cite{Hofman} of the 
formality result proved in \cite{Kontsevich_def_q} for the case 
of the so-called C model of Cattaneo and Felder \cite{CF}. 
Finally, I believe that a similar analysis could be carried out (at the 
2-category level !) for the topological membrane theory formulated 
in \cite{JS}. The relation between the topological membrane of \cite{JS} 
and topological strings seems to be that the first quantization of the 
membrane gives the {\em second} quantization of strings. In particular, it 
is natural to expect that, at least for the topological case,  
an extension of the work of \cite{JS} may
provide a description of `second quantized' D-brane dynamics in terms of the 
first quantized dynamics of topological membranes. In this sense, M-theory may 
amount to a description of second quantized open-closed string dynamics through
the first quantization of a membrane\footnote{And this may suffice, if 
the poorly understood statement that `M-theory has no coupling constant'
is taken to mean that first quantization of membranes somehow
suffices to give a complete description of M-theory.} .

\

{\bf Acknowledgments}
I wish to thank Prof. Sorin Popescu for collaboration in 
a related project and Professor M.~Rocek for 
support and interest in my work. I also thank Prof. M. Rocek, 
J.~S.~Park, R.~Roiban and Y.~Chepelev for interesting comments and 
observations. 
The author is supported by the Research Foundation under NSF grant 
PHY-9722101. 

\


\begin{thebibliography}{100}
\bibitem{com1}{C.~I.~Lazaroiu,
{\em Generalized complexes and string field theory}, hep-th/0102122.
}
\bibitem{tachyon}{A.~Sen, B.~Zwiebach, {\em Tachyon condensation 
in string field theory},  JHEP 0003 (2000) 002, hep-th/9912249; 
L.~Rastelli, A.~Sen, B.~Zwiebach, 
{\em String Field Theory Around the Tachyon Vacuum}, hep-th/0012251; 
 J.~A.~Minahan, B.~Zwiebach, {\em Effective Tachyon Dynamics in 
Superstring Theory}, hep-th/0009246; 
N.~Moeller, A.~Sen, B.~Zwiebach, 
{\em D-branes as Tachyon Lumps in String Field Theory}, 
JHEP 0008 (2000) 039;  N.~Berkovits, A.~Sen, B.~Zwiebach,
{\em Tachyon Condensation in Superstring Field Theory}; 
 W.~Taylor, 
{\em  D-brane effective field theory from string field theory}, 
Nucl.Phys. B585 (2000) 171-192;  
D.~Kutasov, M.~Marino, G.~Moore, 
{\em Some Exact Results on Tachyon Condensation in String Field Theory}, 
hep-th/0009148 ;  D.~Kutasov, M.~Marino, G.~Moore, 
{\em Remarks on Tachyon Condensation in Superstring Field Theory}, 
hep-th/0010108 ; A.~A.~Gerasimov, S.~L.~Shatashvili, 
{\em  On Exact Tachyon Potential in Open String Field Theory}, 
JHEP 0010 (2000) 034.}
\bibitem{MacLane}{ S.~Mac~Lane, 
{\em Categories for the working mathematician},  Graduate texts in mathematics, {\bf 5}, 
New York, Springer, 1971.}
\bibitem{helices}{A.~N.~Rudakov et al, 
{\em Helices and vector bundles} 
(translated by A.~D.~King, P.~Kobak and A.~Maciocia), 
London Mathematical Society lecture note series {\bf 148},      
Cambridge University Press, 1990.}
\bibitem{Keller_dg}{B.~Keller, {\em Deriving DG categories}, 
Ann. Scient. Ecole Normale Sup, $4^e$ serie, t. {\bf 27} (1994) , pp 63-102.}
\bibitem{Bondal_Kapranov}{A.~Bondal, M.~M.~Kapranov, 
{\em Enhanced triangulated categories}, Mat. Sb. {\bf 181} (1990), No.5, 669, 
English translation in Math. USSR Sbornik Vol {\bf 70} (1991), No. 1 , 93. }
\bibitem{Witten_K}{ E.~Witten, 
{\em D-branes and K-theory}, JHEP 9812:019, 1998.}
\bibitem{Witten_K_review}{ E.~Witten, 
{\em Overview Of K-Theory Applied To Strings}, hep-th/0007175.}
\bibitem{Seidel}{P.~Seidel, 
{\em Graded Lagrangian submanifolds},In: Northern California Symplectic 
Geometry Seminar, AMS Translations vol. 196, 1999, pp. 237-250,  
math.SG/9903049.}
\bibitem{Keller}{ B.~Keller, 
{\em Introduction to A-infinity algebras and modules}, 
math.RA/9910179 .}
\bibitem{pi_stab}{ M.~R. Douglas, B.~Fiol, C.~Romelsberger, 
{\em Stability and BPS branes}, hep-th/0002037.}
\bibitem{Oz_triples}{Y.~Oz, T.~Pantev, D.~Waldram,
{\em Brane-Antibrane Systems on Calabi-Yau Spaces}, 
hep-th/0009112.}
\bibitem{Witten_SFT}{E.~Witten, {\em Noncommutative geometry and string 
field theory}, Nucl. Phys, {\bf B268} (1986) 253.}
\bibitem{top}{C.~I.~Lazaroiu, 
{\em On the structure of open-closed topological field 
theory in two dimensions}, hep-th/0010269.}
\bibitem{boundary}{ C.~I.~Lazaroiu, 
{\em Instanton amplitudes in open-closed topological string theory},
hep-th/0011257.}
\bibitem{Hofman}{C.~Hofman, W.~K.~Ma, 
{\em Deformations of topological open strings}, hep-th/006120.}
\bibitem{Kontsevich}{M.~Kontsevich, {\em Homological algebra of mirror 
symmetry},    Proceedings   of    the   International    Congress   of
Mathematicians,     (Zurich,      1994),     120--139,     Birkhauser,
alg-geom/9411018.}
\bibitem{Moore_top}{G.~Moore, 
{\em Some Comments on Branes, G-flux, and K-theory}, 
hep-th/0012007.}
\bibitem{K_bos}{ J.~A.~Harvey, G.~Moore, 
{\em Noncommutative Tachyons and K-Theory}, hep-th/0009030.  }
\bibitem{Douglas_Kontsevich}{M.~Douglas,
{\em D-branes, Categories and N=1 Supersymmetry}, hep-th/0011017.}
\bibitem{Fukaya}{K.~Fukaya, 
{\em  Morse homotopy,  $A^\infty$-category and  Floer  homologies}, in
{\em  Proceedings of  the  GARC Workshop  on  Geometry and  Topology},
ed. by H.~J.~Kim, Seoul  national University (1994), 1-102; {\em Floer
homology, $A^\infty$-categories and topological field theory}, in {\em
Geometry and Physics}, Lecture  notes in pure and applied mathematics,
{\bf 184},  pp 9-32, Dekker, New  York, 1997; {\em  Floer homology and
Mirror       symmetry,      I},       preprint       available      at
$http://www.kusm.kyoto-u.ac.jp/~\tilde{}~fukaya/fukaya.html.$}
\bibitem{Fukaya2}{K. Fukaya, Y.-G. Oh, H.~Ohta, K.~Ono, 
{\em Lagrangian intersection Floer theory - anomaly and obstructon}, 
  preprint       available      at
$http://www.kusm.kyoto-u.ac.jp/~\tilde{}~fukaya/fukaya.html.$}
\bibitem{Witten_CS}{E.~Witten, 
{\em Chern-Simons gauge theory as a string theory}, 
The Floer memorial volume, 637--678, Progr.
Math., 133, Birkhauser, Basel, 1995, hep-th/9207094.}
\bibitem{Witten_NLSM}{E.~Witten, {\em Topological sigma models}, Commun. Math. Phys. {\bf 118} (1988),411.}
\bibitem{Witten_mirror}{E.~Witten, 
{\em Mirror manifolds and topological field theory}, 
In S.T.~Yau, S.T. (ed.): Mirror symmetry I* 121-160,
hep-th/9112056.}
\bibitem{Zwiebach_open}{
B.~Zwiebach,  {\em  Oriented  open-closed  string  theory  revisited},
Annals. Phys. {\bf 267} (1988), 193, hep-th/9705241.}
\bibitem{Gaberdiel}{
M.~Gaberdiel,  B.~Zwiebach, {\em Tensor  constructions of  open string
theories  I:Foundations   },  Nucl.  Phys  {\bf   B505}  (1997),  569,
hep-th/9705038.}
\bibitem{Kontsevich_def_q}{M.~Kontsevich, 
{\em Deformation quantisation of Poisson Manifolds}, I, 
mat/9709010.}
\bibitem{CF}{Alberto S. Cattaneo, Giovanni Felder, 
{\em A path integral approach to the Kontsevich quantization formula}, 
math.QA/9902090.}
\bibitem{com3}{C.~I.~Lazaroiu and S.~Popescu, {\em in preparation}.}
\bibitem{JS}{ Jae-Suk Park, 
{\em Topological open p-branes},  hep-th/0012141.}
\end{thebibliography}
\end{document}